\pdfoutput=1
\documentclass[11pt]{article}

\usepackage[hmargin=1in,vmargin=1in]{geometry}
\usepackage{hchang}
\usepackage[most]{tcolorbox}
\hypersetup{hypertexnames=false}
\nolinenumbers
\usepackage{aliascnt}
\usepackage[nameinlink,capitalize,noabbrev]{cleveref}
\usepackage{float}
\usepackage{pgfplots}
\usepackage{xspace}
\usepackage{etoolbox}
\usetikzlibrary{arrows.meta,backgrounds,calc,fit,shadows.blur,shapes.geometric}
\pgfplotsset{compat=1.18}

\newfloat{paperalgorithm}{tbp}{loa}
\floatname{paperalgorithm}{Algorithm}

\newtheorem{theorem}{Theorem}[section]

\newaliascnt{lemma}{theorem}
\newtheorem{lemma}[lemma]{Lemma}
\aliascntresetthe{lemma}

\newaliascnt{observation}{theorem}
\newtheorem{observation}[observation]{Observation}
\aliascntresetthe{observation}

\newaliascnt{corollary}{theorem}
\newtheorem{corollary}[corollary]{Corollary}
\aliascntresetthe{corollary}

\crefname{corollary}{Corollary}{Corollaries}
\Crefname{corollary}{Corollary}{Corollaries}
\crefname{lemma}{Lemma}{Lemmas}
\Crefname{lemma}{Lemma}{Lemmas}
\crefname{observation}{Observation}{Observations}
\Crefname{observation}{Observation}{Observations}

\newcommand{\OPT}{\operatorname{OPT}}
\newcommand{\TSP}{\textsc{TSP}\xspace}
\newcommand{\MaxTSP}{\textsc{Max-TSP}\xspace}
\newcommand{\eps}{\varepsilon}
\newcommand{\shortedge}{\operatorname{short}}
\newcommand{\longedge}{\operatorname{long}}

\crefname{paperalgorithm}{Algorithm}{Algorithms}
\Crefname{paperalgorithm}{Algorithm}{Algorithms}

\definecolor{figBlue}{HTML}{0068FF}
\definecolor{figOrange}{HTML}{FF5700}
\definecolor{figTeal}{HTML}{00C68A}
\definecolor{figGray}{HTML}{4D5C73}
\definecolor{figInk}{HTML}{16233B}
\definecolor{figOutline}{HTML}{101827}
\definecolor{figPanel}{HTML}{F7FAFF}
\definecolor{figBorder}{HTML}{AFC2DF}
\definecolor{figBlueTint}{HTML}{E1ECFF}
\tikzset{
  soft panel/.style={
    fill=figPanel,
    draw=figBorder,
    line width=0.65pt,
    rounded corners=6pt,
    blur shadow={
      shadow opacity=38,
      shadow blur steps=28,
      shadow blur radius=6.8pt,
      shadow xshift=0.8pt,
      shadow yshift=-3.0pt,
    },
  },
  figure heading/.style={
    font=\small\bfseries,
    text=figInk,
  },
  figure note/.style={
    font=\footnotesize,
    text=figGray,
  },
}
\let\plainbibliography\thebibliography
\let\endplainbibliography\endthebibliography
\renewenvironment{thebibliography}[1]
  {\plainbibliography{#1}%
   \setlength{\itemsep}{0.12\baselineskip}%
   \setlength{\parsep}{0pt}}
  {\endplainbibliography}

\title{Improved Approximation and Hardness Bounds for $\beta$-\TSP}
\author{Paul Liu \and Jack Spalding-Jamieson \and Da Wei Zheng}
\date{}

\begin{document}

\maketitle

\begin{abstract}
The traveling salesman problem (\TSP) asks for a minimum-length Hamiltonian
cycle in a complete weighted graph.  Its approximability depends on the edge
lengths: unrestricted instances admit no constant-factor approximation unless
$\mathrm{P}=\mathrm{NP}$, metric instances admit Christofides'
$3/2$-approximation, and fixed-dimensional Euclidean instances admit a PTAS.
We study $\beta$-\TSP, where the edge lengths satisfy
\[
  \ell_{uw}\leq\beta(\ell_{uv}+\ell_{vw})
  \qquad
  \forall\,u,v,w\in V\ \text{pairwise distinct},
  \quad
  \beta\in\left[\tfrac{1}{2},1\right].
\]
The condition arises naturally in various settings, for example, when metric
travel times are combined with positive service times at the vertices.
The previous best approximation algorithm,
due to B\"ockenhauer et al.,
ranges from $1$ at $\beta=1/2$ to $3/2$ as
$\beta$ approaches $1$.
We obtain a strictly better guarantee for every
$1/2<\beta<1$
by combining two simple constructions whose guarantees counterbalance each other.
We also improve the best known approximation hardness bound for every
$1/2<\beta<1$.
\end{abstract}

\section{Introduction}

The traveling salesman problem (\TSP) asks for a shortest Hamiltonian cycle
in a complete edge-weighted graph.  The approximability of \TSP changes
substantially as additional structure is imposed on the edge lengths.  With
unrestricted lengths, no polynomial-time constant-factor approximation exists
unless $\mathrm{P}=\mathrm{NP}$.  Under the triangle inequality (called \emph{metric} \TSP), Christofides'
algorithm achieves a $3/2$-approximation, while fixed-dimensional Euclidean
\TSP admits a PTAS~\cite{christofides2022worst,arora1998polynomial}.
Stronger
or different structural assumptions yield further improvements: graphic \TSP (the distances are shortest-path distances in an unweighted graph)
has a $7/5$-approximation, and $(1,2)$-\TSP (all distances are either $1$ or $2$)
has an
$8/7$-approximation~\cite{sebo2014shorter,adamaszek2018new}.
Even the ``metric''
asymmetric variant, in which edge lengths are directed, but still satisfy the triangle inequality, admits a constant-factor
approximation~\cite{svensson2020constant}.  We study
$\beta$-\TSP, which interpolates between equal edge lengths and general
metrics.

Let $K_V$ be the complete graph on a finite set $V$, with symmetric,
positive edge lengths $\ell_{uv}$ for $u,v\in V$.  For
$1/2\leq\beta\leq1$, a $\beta$-\TSP instance satisfies
\begin{equation}
  \ell_{uw}\leq\beta(\ell_{uv}+\ell_{vw})
  \qquad\text{for all distinct $u,v,w$.}
  \label{eq:beta-inequality}
\end{equation}
At $\beta=1$ we recover metric \TSP.  At $\beta=1/2$, the three inequalities
on each triangle force all edge lengths to be equal.  Thus any tour is
optimal, giving a trivial $1$-approximation.  Write $\ell(T)$ for the length
of a tour $T$ and $\OPT$ for the optimum length.

The $\beta$-\TSP condition arises naturally in several settings.  For example,
suppose vertices correspond to customers,
and that a route includes time spent at each customer.  Let $d$ be the
travel-time metric and let $s_v>0$ be the service time at
$v$~\cite{tas2016traveling}.  Set
$\ell_{uv}=d(u,v)+(s_u+s_v)/2$.  Every tour then has length
$d(T)+\sum_v s_v$, and
\[
  \ell_{uv}+\ell_{vw}-\ell_{uw}
  \geq s_v>0.
\]
Because $V$ is finite, one value $\beta<1$ works for every triple.  Uniform
service times give the simpler model $\ell_{uv}=d(u,v)+\lambda$.  The same
argument covers every finite strict metric, including Euclidean point sets
with no three collinear.

Bounded metrics provide another natural source of $\beta$-\TSP instances.  If
all edge lengths lie in $[m,M]$ and $M\leq2\beta m$, then
\[
  \ell_{uw}\leq M\leq2\beta m
  \leq\beta(\ell_{uv}+\ell_{vw}),
\]
so \cref{eq:beta-inequality} holds.  In particular, replacing the lengths $1$
and $2$ in a $(1,2)$-\TSP instance by $1$ and $2\beta$, respectively, produces
a $\beta$-\TSP instance.  This observation links $\beta$-\TSP to work on
$(1,2)$-\TSP and bounded
 metrics~\cite{papadimitriou1993traveling,engebretsen2001approximation,
  karpinski2012approximation}.  In particular, Karpinski and
  Schmied~\cite{karpinski2012approximation} show that $(1,2)$-\TSP is
  inapproximable within $535/534$, which implies a
  $1+(2\beta-1)/534$ lower bound for $\beta$-\TSP.

Related sharpenings of the triangle inequality
have also appeared in
linear-time approximation~\cite{bilo2011approximating}, asymmetric
\TSP~\cite{chandran2002approximations,blaeser2006improved,kowalik2009two},
multiobjective \TSP~\cite{manthey2009approximation}, and
reoptimization~\cite{dandrea2018stability}.

\subsection{Our results}

B\"ockenhauer, Hromkovi\v{c}, Klasing, Seibert, and Unger introduced
$\beta$-\TSP~\cite{bockenhauer2000approximation}.  Their approximation factor is
\[
  \min\left\{
    \frac{2-\beta}{3(1-\beta)},
    \frac{3\beta^2}{3\beta^2-2\beta+1}
  \right\}.
\]
No better general bound has been proved since, but several hardness results
followed~\cite{bockenhauer2000improved,
engebretsen2001approximation,engebretsen2006tsp,karpinski2012approximation}.

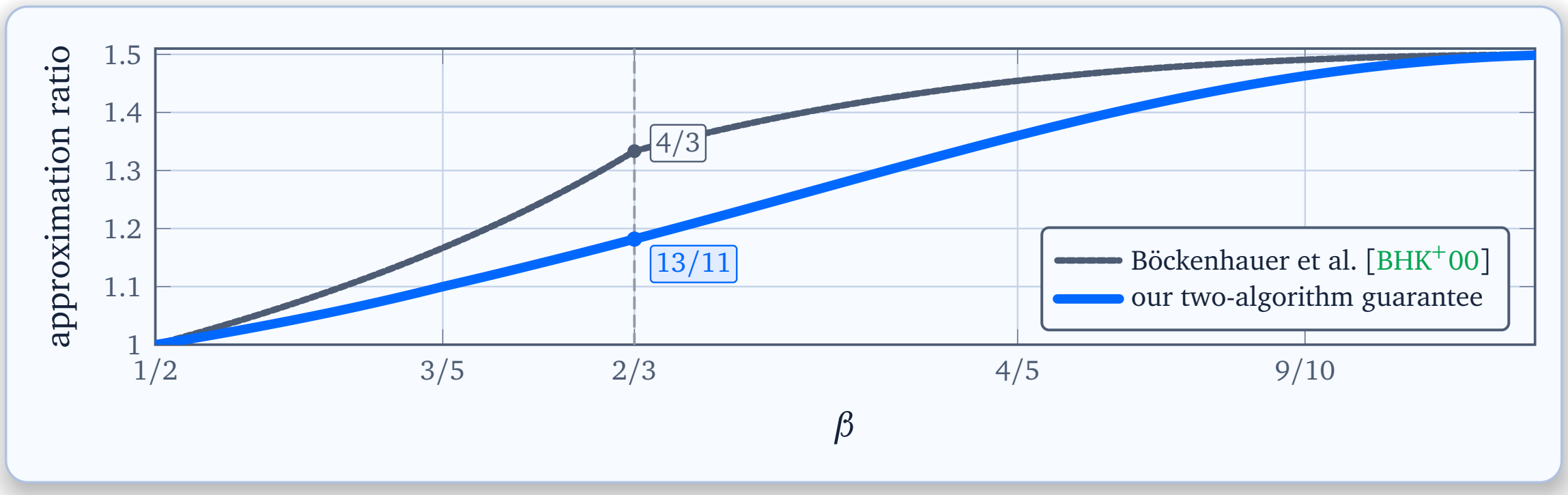
\begin{figure}[H]
  \centering
  \begin{tikzpicture}
    \begin{axis}[
      width=14.95cm,
      height=4.45cm,
      xmin=0.5,
      xmax=0.98,
      ymin=1,
      ymax=1.51,
      xlabel={$\beta$},
      ylabel={approximation ratio},
      xtick={0.5,0.6,0.6666667,0.8,0.9},
      xticklabels={$1/2$,$3/5$,$2/3$,$4/5$,$9/10$},
      ytick={1,1.1,1.2,1.3,1.4,1.5},
      axis line style={draw=figGray,line width=0.7pt},
      tick style={draw=figGray},
      tick label style={font=\scriptsize,text=figGray},
      label style={font=\small,text=figInk},
      grid=major,
      grid style={draw=figBorder!72,line width=0.45pt},
      axis background/.style={fill=figPanel},
      legend style={
        at={(0.982,0.045)},
        anchor=south east,
        draw=figGray,
        line width=0.75pt,
        rounded corners=2pt,
        fill=figPanel,
        font=\scriptsize,
        text=figInk,
        inner sep=3pt,
        row sep=-1pt,
      },
      legend cell align={left},
      samples=280,
      every axis plot/.append style={line cap=round,line join=round},
    ]
      \addplot[
        figGray,
        densely dotted,
        line width=1.8pt,
        domain=0.5:0.98,
      ]
      {min((2-x)/(3*(1-x)),(3*x^2)/(3*x^2-2*x+1))};
      \addlegendentry{B\"ockenhauer et al.~\cite{bockenhauer2000approximation}}

      \addplot[
        figBlue,
        line width=2.6pt,
        domain=0.5:0.6,
      ]
      {(4-3*x)/(5*(1-x))};
      \addplot[
        figBlue,
        line width=2.6pt,
        domain=0.6:0.98,
      ]
      {(8*x^2-13*x+8)/(2*(5*x^2-9*x+5))};
      \addlegendimage{figBlue,line width=2.6pt,line cap=round}
      \addlegendentry{our two-algorithm guarantee}

      \draw[figGray!62,densely dashed,line width=0.72pt,line cap=round]
        (axis cs:0.6666667,1)--(axis cs:0.6666667,1.51);

      \fill[figGray] (axis cs:0.6666667,1.3333333) circle[radius=1.9pt];
      \fill[figBlue] (axis cs:0.6666667,1.1818182) circle[radius=2.1pt];

      \node[
        anchor=west,font=\scriptsize\bfseries,text=figGray,
        fill=figPanel,draw=figGray,line width=0.45pt,
        rounded corners=1pt,inner sep=1.4pt,
      ] at (axis cs:0.672,1.347) {$4/3$};
      \node[
        anchor=north west,font=\scriptsize\bfseries,text=figBlue,
        fill=figBlueTint,draw=figBlue,line width=0.45pt,
        rounded corners=1pt,inner sep=1.4pt,
      ] at (axis cs:0.672,1.172) {$13/11$};
    \end{axis}
    \coordinate (plot south west) at (current bounding box.south west);
    \coordinate (plot north east) at (current bounding box.north east);
    \begin{scope}[on background layer]
      \node[soft panel,fit=(plot south west)(plot north east),inner sep=7pt] {};
    \end{scope}
  \end{tikzpicture}
  \caption{
    Approximation guarantees as a function of $\beta$.
  }
  \label{fig:upper-curves}
\end{figure}

To improve this bound, which has stood for more than two decades, we combine
two simple algorithms whose guarantees counterbalance one another as $\beta$
varies.
Our algorithm returns the shorter of two tours:
A Christofides tour over shifted weights,
and a tour obtained through a reduction to \MaxTSP.
Both analyses use
the statistic $S=\sum_v\min_{u\ne v}\ell_{uv}$.  The first bound
improves as $S$ grows; the second improves as $S$ shrinks.  Carefully
counterbalancing these two tours gives
\begin{equation*}
  \begin{cases}
    \dfrac{4-3\beta}{5(1-\beta)},
      &\dfrac12\leq\beta\leq\dfrac35,\\[6pt]
    \dfrac{8(1-\beta)^2+3\beta}
          {2\left(5(1-\beta)^2+\beta\right)},
      &\dfrac35\leq\beta<1.
  \end{cases}
\end{equation*}
This is a significant improvement, as can be seen in \cref{fig:upper-curves}.

We also prove a new lower bound for $(1,2)$-\TSP, which in turn yields a better
lower bound for $\beta$-\TSP.
\begin{theorem}
\label{thm:one-two-hardness}
It is NP-hard to approximate $(1,2)$-\TSP better than $2639/2634$.
\end{theorem}
The previous inapproximability bound of Karpinski and
Schmied~\cite{karpinski2012approximation} was $535/534=1+1/534$, whereas
our excess is $5/2634>1/534$.
This also gives an improved lower bound for $\beta$-\TSP.
\begin{corollary}
\label{thm:hardness}
For every fixed $1/2<\beta<1$, it is NP-hard to approximate $\beta$-\TSP
within any factor smaller than
\[ 1+\frac{5(2\beta-1)}{2634}.  \]
\end{corollary}

\section{Two Complementary Algorithms}

Our algorithm constructs two candidate tours using different transformations
of the edge lengths.  One transformation allows us to apply Christofides'
$\frac32$-approximation~\cite{christofides2022worst}; the other reduces the
problem to \MaxTSP, for which we use a $4/5$-approximation~\cite{dudycz20174}.
We return the shorter candidate, allowing the two guarantees to compensate for
each other's weaker cases.
Our full algorithm can be found in
\cref{alg:two-tours}.

Throughout this section,
we denote
$\shortedge(v):=\min_{x\neq v}\ell_{vx}$
and
$\longedge(v):=\max_{x\neq v}\ell_{vx}$.

\begin{paperalgorithm}[H]
\begin{tcolorbox}[
  enhanced,
  colback=figPanel,
  colframe=figBorder,
  boxrule=0.65pt,
  arc=5pt,
  left=8pt,
  right=8pt,
  top=7pt,
  bottom=6pt,
  borderline west={3pt}{0pt}{figBlue},
  shadow={0.8mm}{-1.2mm}{0mm}{figOutline!20},
]
\small
\textit{Input.}
A complete graph with edge lengths $\ell$, and a value $\beta$.

\begin{enumerate}[label=\textbf{\arabic*.},leftmargin=2.1em,itemsep=2pt,topsep=4pt]
  \item Subtract $(1-\beta)(\shortedge(u)+\shortedge(v))$ from every
    edge $uv$.
    Run Christofides' algorithm on the new lengths to obtain
    $T_{\rm shift}$.
  \item For each edge $uv$, define
    \[
      w_{uv}:=\frac{\longedge(u)+\longedge(v)}2-\ell_{uv}.
    \]
    Apply a $4/5$-approximation for \MaxTSP using the weights $w$, and denote
    the returned tour by $T_{\rm max}$.
  \item Return whichever of $T_{\rm shift}$ and $T_{\rm max}$ has smaller
    length under the original edge lengths $\ell$.
\end{enumerate}
\end{tcolorbox}
\caption{The approximation algorithm for $\beta$-\TSP.}
\label{alg:two-tours}
\end{paperalgorithm}

\begin{theorem}
\label{thm:upper}
Let $T$ be the tour returned by \cref{alg:two-tours} on a
$\beta$-\TSP instance, where $1/2\leq\beta<1$.
Then
\[
  \ell(T)\leq
  \OPT\cdot
  \begin{cases}
    \dfrac{4-3\beta}{5(1-\beta)},
      &\dfrac12\leq\beta\leq\dfrac35,\\[6pt]
    \dfrac{8(1-\beta)^2+3\beta}
          {2\left(5(1-\beta)^2+\beta\right)},
      &\dfrac35\leq\beta<1.
  \end{cases}.
\]
\end{theorem}

We first bound $\ell(T_{\rm shift})$ and $\ell(T_{\rm max})$ separately, and
then combine the two estimates.

\subsection{The two approximations}

We start with some helpful observations.

\begin{observation}
\label{obs:edge-range-1}
For every vertex $v$,
\[
  \longedge(v)
  \leq
  \frac{\beta}{1-\beta}\shortedge(v).
\]
\end{observation}

\begin{proof}
Choose vertices $x,z$ realizing $\shortedge(v)=\ell_{xv}$ and
$\longedge(v)=\ell_{vz}$.  If $x=z$, the claim follows from
$\beta/(1-\beta)\geq1$.  Otherwise, two applications of
\cref{eq:beta-inequality} give
\[
  \longedge(v)
  \leq\beta\bigl(\shortedge(v)+\ell_{xz}\bigr)
  \leq\beta\bigl((1+\beta)\shortedge(v)
                  +\beta\longedge(v)\bigr).
\]
Rearranging and using $1-\beta^2=(1-\beta)(1+\beta)$ proves the result.
\end{proof}

\begin{observation}
\label{obs:edge-range-2}
For every $\TSP$ instance,
\[
  \sum_v\shortedge(v)
  \leq \OPT
  \leq \sum_v\longedge(v).
\]
\end{observation}

\begin{proof}
Let $T$ be an optimum tour.  At every vertex $v$, both edges of $T$ incident
to $v$ have length at least $\shortedge(v)$.  Summing over the vertices counts
every tour edge twice, so
\[
  2\OPT\geq2\sum_v\shortedge(v).
\]
For the upper bound, orient $T$ arbitrarily and charge each edge to its tail.
Each vertex is charged once, by an edge of length at most $\longedge(v)$, and
therefore
\[
  \OPT\leq\sum_v\longedge(v).
\]
\end{proof}

We now prove a result for the ``shifted'' Christofides step.

\begin{lemma}
\label{lem:shift-tour}
For every $\beta$-\TSP instance with $1/2\leq\beta<1$, the tour
$T_{\rm shift}$ produced in Step~1 of \cref{alg:two-tours} satisfies
\[
  \ell(T_{\rm shift})
  \leq
  \frac32\OPT-(1-\beta)\sum_v\shortedge(v).
\]
\end{lemma}

\begin{proof}
Write $\ell'$ for the new lengths.
These lengths are nonnegative since
\[
  \ell'_{uv}
  \geq \ell_{uv}-2(1-\beta)\ell_{uv}
  =(2\beta-1)\ell_{uv}\geq0.
\]
They satisfy the triangle inequality because
\[
\begin{aligned}
  \ell'_{uv}+\ell'_{vw}-\ell'_{uw}
  &=
  \ell_{uv}+\ell_{vw}-\ell_{uw}
  -2(1-\beta)\shortedge(v)\\
  &\geq
  (1-\beta)(\ell_{uv}+\ell_{vw})
  -2(1-\beta)\shortedge(v)\\
  &=
  (1-\beta)
  \bigl(\ell_{uv}+\ell_{vw}-2\shortedge(v)\bigr)\\
  &\geq0.
\end{aligned}
\]
Every tour loses length exactly
$2(1-\beta)\sum_v\shortedge(v)$, since each vertex belongs to two tour
edges.
This is a constant independent of the tour itself.
The optimum under $\ell'$ is therefore
$\OPT-2(1-\beta)\sum_v\shortedge(v)$, and Christofides' guarantee gives
\[
\begin{aligned}
  \ell(T_{\rm shift})
  &=
  \ell'(T_{\rm shift})
  +2(1-\beta)\sum_v\shortedge(v)\\
  &\leq
  \frac32\left(
    \OPT-2(1-\beta)\sum_v\shortedge(v)
  \right)
  +2(1-\beta)\sum_v\shortedge(v)\\
  &=\frac32\OPT-(1-\beta)\sum_v\shortedge(v).
\end{aligned}
\]
\end{proof}

\begin{lemma}
\label{lem:max-tour}
For every $\beta$-\TSP instance with $1/2\leq\beta<1$, the tour
$T_{\rm max}$ produced in Step~2 of \cref{alg:two-tours} satisfies
\[
  \ell(T_{\rm max})
  \leq
  \frac45\OPT
  +\frac{\beta}{5(1-\beta)}\sum_v\shortedge(v)
  \leq
  \frac{4-3\beta}{5(1-\beta)}\OPT.
\]
\end{lemma}

\begin{proof}
The weights are nonnegative since $\ell_{uv}\leq\min\left\{\longedge(u),\longedge(v)\right\}$.
Every tour $T$ satisfies
\[
  w(T)=\sum_v\longedge(v)-\ell(T).
\]
The maximum $w$-weight of a tour is therefore $\sum_v\longedge(v)-\OPT$,
and
the $4/5$-approximation therefore guarantees
\[
  w(T_{\rm max})
  \geq
  \frac45\left(\sum_v\longedge(v)-\OPT\right).
\]
Hence, by applying \cref{obs:edge-range-1,obs:edge-range-2},
\[
\begin{aligned}
  \ell(T_{\rm max})
  &=
  \sum_v\longedge(v)-w(T_{\rm max})\\
  &\leq
  \sum_v\longedge(v)
  -\frac45\left(\sum_v\longedge(v)-\OPT\right)\\
  &=
  \frac45\OPT+\frac15\sum_v\longedge(v)\\
  &\leq
  \frac45\OPT
  +\frac{\beta}{5(1-\beta)}\sum_v\shortedge(v)\\
  &\leq
  \left(
    \frac45+\frac{\beta}{5(1-\beta)}
  \right)\OPT\\
  &=
  \frac{4-3\beta}{5(1-\beta)}\OPT.
\end{aligned}
\]
\end{proof}

\subsection{Balancing the estimates}

\begin{proofof}{\cref{thm:upper}}
Since $T$ is no longer than $T_{\rm max}$,
\cref{lem:max-tour} gives
\begin{equation}
\label{eq:direct-max-bound}
  \ell(T)
  \leq
  \frac{4-3\beta}{5(1-\beta)}\OPT.
\end{equation}

Next,
we will apply the bounds involving $\sum_v\shortedge(v)$
in
\cref{lem:shift-tour,lem:max-tour}
using a careful linear combination so that the terms involving $\sum_v\shortedge(v)$
cancel.
That is,
\[
\begin{aligned}
  \bigl(\beta+5(1-\beta)^2\bigr)\ell(T)
  &\leq
  \beta\ell(T_{\rm shift})
  +5(1-\beta)^2\ell(T_{\rm max})\\
  &\leq
  \beta\left(
    \frac32\OPT-(1-\beta)\sum_v\shortedge(v)
  \right)\\
  &\qquad
  +5(1-\beta)^2\left(
    \frac45\OPT
    +\frac{\beta}{5(1-\beta)}
      \sum_v\shortedge(v)
  \right)\\
  &=
  \left(\frac{3\beta}{2}+4(1-\beta)^2\right)\OPT\\
  &\qquad
  +\bigl(-\beta(1-\beta)+\beta(1-\beta)\bigr)
    \sum_v\shortedge(v)\\
  &=
  \frac12\left(3\beta+8(1-\beta)^2\right)\OPT.
\end{aligned}
\]
Therefore,
\begin{equation}
\label{eq:balanced-max-bound}
  \ell(T)
  \leq
  \frac{8(1-\beta)^2+3\beta}
       {2\bigl(5(1-\beta)^2+\beta\bigr)}\OPT.
\end{equation}

We briefly show that we have made the better choice of the two bounds
for every value of $\beta$.
The factor in \cref{eq:direct-max-bound} minus that in
\cref{eq:balanced-max-bound} is
\[
  \frac{\beta(2\beta-1)(5\beta-3)}
       {10(1-\beta)\bigl(5(1-\beta)^2+\beta\bigr)}.
\]
All factors other than $5\beta-3$ are nonnegative, and the denominator is
positive.
Thus the bound in \cref{eq:direct-max-bound} is better when
$\beta\leq3/5$, while the bound in \cref{eq:balanced-max-bound} is better
when $\beta\geq3/5$.
\end{proofof}

\section{Hardness}
\label{sec:hardness}

In this section we give an overview of the proof of \cref{thm:one-two-hardness}.  The $\beta$-\TSP lower
bound in \cref{thm:hardness} will then follow by rescaling the weight-$2$
edges.  Recall that a $(1,2)$-\TSP instance is a complete graph in which every
edge has length one or two.  A tour on $N$ vertices uses exactly $N$ edges, so
its length is $N+q$, where $q$ is the number of length-two edges in the tour.
Hardness reductions exploit this identity by arranging that $q$ records how
many Boolean constraints are violated.

We start from H{\aa}stad's seminal PCP hardness result for MAX-E3-LIN-2,
whose constraints specify whether the sum of three Boolean variables is even
or odd~\cite[Theorem~5.4]{hastad2001optimal}.  For every fixed $\delta>0$, it
is NP-hard to distinguish instances in which some assignment satisfies at
least a $1-\delta$ fraction of the equations from instances in which no
assignment satisfies more than a $1/2+\delta$ fraction.  We use the
fixed-occurrence refinement of Chleb\'ik and
Chleb\'ikov\'a~\cite[Theorem~2]{chlebik2003approximation}, in which every
variable occurs the same fixed number of times.

The reduction then turns these equations into a $(1,2)$-\TSP instance.  The
two possible values of a Boolean variable are represented by two standard ways
of traversing a constant-size gadget.  The equation gadgets convert violated
equations into additional length-two edges in the tour.  Different occurrences
of one original variable are initially represented by separate copies.  To
enforce consistency, we use the amplifier of Chleb\'ik and
Chleb\'ikov\'a~\cite[Section~3 and Theorem~4]{chlebik2003approximation}.  It
connects the copies with equality equations.  Its expansion property ensures
that many disagreeing copies produce many violated equalities.  The copies can
therefore be rounded to their majority value without substantially changing
the number of violations.

The Chleb\'ik--Chleb\'ikov\'a amplifier contains two cycles, but the
Karpinski--Schmied reduction~\cite{karpinski2012approximation} expects the
consistency equations for each original variable to form one circle of equation systems.
We splice the two cycles into one.
This splice loses
at most two consistency violations per amplifier to have the form required by Karpinski and Schmied.  Their construction
translates the number of violated equations into the amount by which the tour
length exceeds its baseline value $N$.  A complete exposition of the construction and its accounting are given in \cref{app:one-two-gap}.

\begin{lemma}
\label{lem:one-two-gap}
For every $0<\xi<5/5268$, it is NP-hard, given a complete $(1,2)$-\TSP
instance $G$ on $N$ vertices, to distinguish
\[
  \OPT(G)\leq(1+\xi)(N-1)+1
  \qquad\text{from}\qquad
  \OPT(G)\geq
  \left(1+\frac5{2634}-\xi\right)(N-1)+1.
\]
\end{lemma}

The proof of \cref{thm:one-two-hardness} follows from the above lemma and the following observation.

\begin{observation}[Gap-to-approximation]
\label{obs:gap-to-approximation}
Let $y,g>0$.  Suppose that, for every sufficiently small $\xi>0$, it is
NP-hard, on instances with an unbounded scale parameter $m$, to distinguish
\[
  \OPT\leq(y+\xi)m+O(1)
  \qquad\text{from}\qquad
  \OPT\geq(y+g-\xi)m-O(1).
\]
The constants hidden in $O(1)$ may depend on the fixed parameters, but not on
$m$.  Then, for every $\eps>0$, it is NP-hard to approximate within
$1+g/y-\eps$.
\end{observation}

\begin{proof}
Choose $\xi$ so that $(y+g-\xi)/(y+\xi)>1+g/y-\eps$, and then choose $m$
sufficiently large to absorb the additive constants.
\end{proof}

\section{Acknowledgements}

Our results were devised with aid from GPT 5.6, specifically in the choice of hardness gadgets and handling of inequalities across sections 2 and 3. The writing in this paper was also
aided by the same model, although the final document is human-generated, and we take full responsibility for the results.

{\small
\bibliographystyle{alpha}
\bibliography{ref}

\newcommand{\etalchar}[1]{$^{#1}$}
\begin{thebibliography}{DMPR17}

\bibitem[AMP18]{adamaszek2018new}
Anna Adamaszek, Matthias Mnich, and Katarzyna Paluch.
\newblock New approximation algorithms for $(1,2)$-{TSP}.
\newblock In {\em 45th International Colloquium on Automata, Languages, and
  Programming}, volume 107 of {\em Leibniz International Proceedings in
  Informatics}, pages 9:1--9:14, 2018.

\bibitem[Aro98]{arora1998polynomial}
Sanjeev Arora.
\newblock Polynomial time approximation schemes for euclidean traveling
  salesman and other geometric problems.
\newblock {\em Journal of the ACM}, 45(5):753--782, 1998.

\bibitem[BFP11]{bilo2011approximating}
Davide Bil{\`o}, Luca Forlizzi, and Guido Proietti.
\newblock Approximating the metric tsp in linear time.
\newblock {\em Theory of Computing Systems}, 49(3):615--631, 2011.

\bibitem[BHK{\etalchar{+}}00]{bockenhauer2000approximation}
Hans-Joachim B{\"o}ckenhauer, Juraj Hromkovi{\v{c}}, Ralf Klasing, Sebastian
  Seibert, and Walter Unger.
\newblock Approximation algorithms for the tsp with sharpened triangle
  inequality.
\newblock {\em Information Processing Letters}, 75(3):133--138, 2000.

\bibitem[BMS06]{blaeser2006improved}
Markus Bl{\"a}ser, Bodo Manthey, and Ji{\v{r}}{\'\i} Sgall.
\newblock An improved approximation algorithm for the asymmetric tsp with
  strengthened triangle inequality.
\newblock {\em Journal of discrete algorithms}, 4(4):623--632, 2006.

\bibitem[BS00]{bockenhauer2000improved}
Hans-Joachim B{\"o}ckenhauer and Sebastian Seibert.
\newblock Improved lower bounds on the approximability of the traveling
  salesman problem.
\newblock {\em RAIRO-Theoretical Informatics and Applications}, 34(3):213--255,
  2000.

\bibitem[CC03]{chlebik2003approximation}
Miroslav Chleb{\'\i}k and Janka Chleb{\'\i}kov{\'a}.
\newblock Approximation hardness for small occurrence instances of np-hard
  problems.
\newblock In {\em Italian Conference on Algorithms and Complexity}, pages
  152--164. Springer, 2003.

\bibitem[Chr22]{christofides2022worst}
Nicos Christofides.
\newblock Worst-case analysis of a new heuristic for the travelling salesman
  problem.
\newblock {\em Operations Research Forum}, 3(1):20, 2022.

\bibitem[CR02]{chandran2002approximations}
L.~Sunil Chandran and L.~Shankar Ram.
\newblock Approximations for {ATSP} with parametrized triangle inequality.
\newblock In {\em Proceedings of the 19th Annual Symposium on Theoretical
  Aspects of Computer Science}, volume 2285 of {\em Lecture Notes in Computer
  Science}, pages 227--237, 2002.

\bibitem[DFP18]{dandrea2018stability}
Annalisa D’Andrea, Luca Forlizzi, and Guido Proietti.
\newblock Stability of reapproximation algorithms for the $\beta$-metric
  traveling salesman (path) problem.
\newblock In {\em Adventures Between Lower Bounds and Higher Altitudes: Essays
  Dedicated to Juraj Hromkovi{\v{c}} on the Occasion of His 60th Birthday},
  pages 156--171. Springer, 2018.

\bibitem[DMPR17]{dudycz20174}
Szymon Dudycz, Jan Marcinkowski, Katarzyna Paluch, and Bartosz Rybicki.
\newblock A 4/5-approximation algorithm for the maximum traveling salesman
  problem.
\newblock In {\em International Conference on Integer Programming and
  Combinatorial Optimization}, pages 173--185. Springer, 2017.

\bibitem[EK01]{engebretsen2001approximation}
Lars Engebretsen and Marek Karpinski.
\newblock Approximation hardness of tsp with bounded metrics.
\newblock In {\em International Colloquium on Automata, Languages, and
  Programming}, pages 201--212. Springer, 2001.

\bibitem[EK06]{engebretsen2006tsp}
Lars Engebretsen and Marek Karpinski.
\newblock Tsp with bounded metrics.
\newblock {\em Journal of Computer and System Sciences}, 72(4):509--546, 2006.

\bibitem[Hå01]{hastad2001optimal}
Johan Håstad.
\newblock Some optimal inapproximability results.
\newblock {\em Journal of the ACM}, 48(4):798--859, 2001.

\bibitem[KM09]{kowalik2009two}
{\L}ukasz Kowalik and Marcin Mucha.
\newblock Two approximation algorithms for {ATSP} with strengthened triangle
  inequality.
\newblock In {\em Algorithms and Data Structures: 11th International Symposium,
  {WADS} 2009}, volume 5664 of {\em Lecture Notes in Computer Science}, pages
  471--482, 2009.

\bibitem[KS12]{karpinski2012approximation}
Marek Karpinski and Richard Schmied.
\newblock On approximation lower bounds for tsp with bounded metrics.
\newblock {\em arXiv preprint arXiv:1201.5821}, 2012.

\bibitem[MSR09]{manthey2009approximation}
Bodo Manthey and L~Shankar~Ram.
\newblock Approximation algorithms for multi-criteria traveling salesman
  problems.
\newblock {\em Algorithmica}, 53(1):69--88, 2009.

\bibitem[PY93]{papadimitriou1993traveling}
Christos~H. Papadimitriou and Mihalis Yannakakis.
\newblock The traveling salesman problem with distances one and two.
\newblock {\em Mathematics of Operations Research}, 18(1):1--11, 1993.

\bibitem[STV20]{svensson2020constant}
Ola Svensson, Jakub Tarnawski, and L{\'a}szl{\'o}~A. V{\'e}gh.
\newblock A constant-factor approximation algorithm for the asymmetric
  traveling salesman problem.
\newblock {\em J. ACM}, 67(6), November 2020.

\bibitem[SV14]{sebo2014shorter}
Andr{\'a}s Seb{\H{o}} and Jens Vygen.
\newblock Shorter tours by nicer ears: $7/5$-approximation for the graph-{TSP},
  $3/2$ for the path version, and $4/3$ for two-edge-connected subgraphs.
\newblock {\em Combinatorica}, 34(5):597--629, 2014.

\bibitem[TGJL16]{tas2016traveling}
Duygu Ta{\c{s}}, Michel Gendreau, Ola Jabali, and Gilbert Laporte.
\newblock The traveling salesman problem with time-dependent service times.
\newblock {\em European Journal of Operational Research}, 248(2):372--383,
  2016.

\end{thebibliography}
}

\clearpage
\appendix
\section{Proof of the \texorpdfstring{$(1,2)$}{(1,2)}-\TSP gap}
\label{app:one-two-gap}

This appendix proves \cref{lem:one-two-gap}.  We give the amplifier and splice
arguments in full and state the precise interface used from the
Karpinski--Schmied reduction.

\paragraph{The original equation gap.}
An instance of MAX-E3-LIN-2 consists of equations
\[
  x_i\oplus x_j\oplus x_k=b
\]
over $\{0,1\}$.  Let $E$ denote the number of equations.  We use the
fixed-occurrence hardness theorem of Chleb\'ik and
Chleb\'ikov\'a~\cite[Theorem~2]{chlebik2003approximation}, based on H{\aa}stad's
parity-gap theorem~\cite[Theorem~5.4]{hastad2001optimal}.  For every fixed
$0<\delta<1/4$ and every sufficiently large constant $d$, it is NP-hard to
distinguish instances in which some assignment violates at most $\delta E$
equations from instances in which every assignment violates at least
$(1/2-\delta)E$ equations, even when every variable occurs exactly $d$ times.
We call these three-variable equations the \emph{original equations} to
distinguish them from the equality equations introduced next.

\paragraph{Amplifying consistency.}
Fix an original variable $x$.  Replace its $d$ occurrences by distinct Boolean
variables indexed by a set $D_x$ of $d$ \emph{contacts}, using one contact in
each original equation.  The contacts are vertices of an amplifier graph
$A_x$; its other vertices are called \emph{checkers}.  Associate a Boolean
variable $y_v$ with every $v\in V(A_x)$ and replace every amplifier edge $uv$
by the equality equation
\[
  y_u\oplus y_v=0.
\]
Without these equations, the occurrence copies could choose their values
independently and the original gap would disappear.

The amplifier property is that, for every $S\subseteq V(A_x)$,
\[
  |\delta_{A_x}(S)|
  \geq\min\{|S\cap D_x|,|D_x\setminus S|\}.
\]
Here $\delta_{A_x}(S)$ is the set of amplifier edges with exactly one endpoint
in $S$.  To see what the inequality accomplishes, fix any assignment and let
$b$ be a majority value among the contacts.  Put
$S=\{v:y_v\neq b\}$ and write $q=|S\cap D_x|$; then $q\leq d/2$.  Every edge
of $\delta_{A_x}(S)$ is a violated equality equation, so at least $q$
equalities are violated.  Rounding all variables of $A_x$ to $b$ repairs those
equalities and can change only the $q$ original equations incident to the
minority contacts.  Thus all copies of $x$ can be made consistent without
increasing the total number of violations.

For every $\eta>0$, Chleb\'ik and Chleb\'ikov\'a construct such amplifiers for
arbitrarily large $d$, with $d$ contacts and $\rho d$ vertices, where
\[
  \rho\leq\frac{69}{10}+\eta
\]
\cite[Section~3 and Theorem~4]{chlebik2003approximation}.  Their amplifier is
the union of two vertex-disjoint cycles covering all contacts and checkers and
an edge-disjoint perfect matching on the checkers.  Consequently, each contact
appears in two equality equations and one original equation, while each checker
appears in three equality equations.  The resulting system is therefore a
\emph{Hybrid instance}: every variable appears exactly three times.

\paragraph{Matching the Karpinski--Schmied interface.}
Karpinski and Schmied organize the consistency equations for each original
variable into a \emph{circle}~\cite[Section~3.1]{karpinski2012approximation}.
A circle is a component of the Hybrid equation system, not a tour in the final
\TSP instance: its circle equations form one cycle through the contacts and
checkers, and its matching equations form a perfect matching on the checkers.
The amplifier above has the required matching but has two cycles rather than
one.  Consecutive contacts are separated by at least five checkers, so each
cycle contains a checker--checker edge and the following splice applies.

\begin{lemma}[Wheel splice]
\label{lem:wheel-splice}
Let two vertex-disjoint cycles contain contact and checker vertices, and let an
edge-disjoint perfect matching join the checker vertices.  Suppose that each
cycle contains a checker--checker edge.  One can replace the two cycles by one
cycle so that
\begin{enumerate}
  \item no matching edge is used by the new cycle;
  \item the two cycle neighbors of every contact are unchanged; and
  \item for every vertex set $S$, the number of cycle edges in $\delta(S)$
        decreases by at most two.
\end{enumerate}
\end{lemma}

\begin{proof}
Delete checker--checker edges $ab$ and $cd$, one from each cycle.  Either
pairing $ac,bd$ or $ad,bc$ closes the two exposed paths into one cycle.  If
neither edge of the first pairing belongs to the matching, use it.  Otherwise,
after relabeling assume that $ac$ is a matching edge.  Matching edges have
disjoint endpoints, so neither $ad$ nor $bc$ is a matching edge and the second
pairing works.  All changed edges have checker endpoints, and only the two
deleted edges can reduce the size of a cut.  This proves the three claims.
\end{proof}

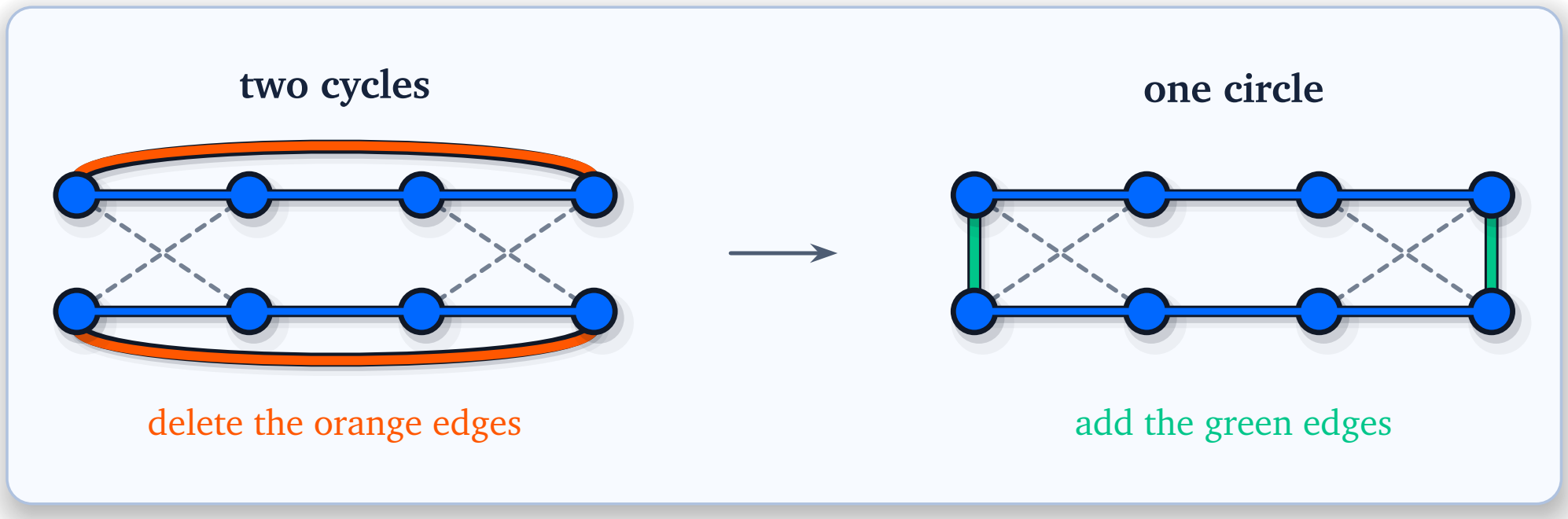
\begin{figure}[H]
  \centering
  \begin{tikzpicture}[
    x=1cm,
    y=0.72cm,
    blue shadow far/.style={
      draw=figOutline,
      opacity=0.05,
      line width=5.2pt,
      line cap=round,
      line join=round,
      transform canvas={xshift=1.0pt,yshift=-1.4pt},
    },
    blue shadow near/.style={
      draw=figOutline,
      opacity=0.15,
      line width=4.0pt,
      line cap=round,
      line join=round,
      transform canvas={xshift=0.55pt,yshift=-0.8pt},
    },
    blue outline/.style={
      draw=figOutline,
      line width=2.75pt,
      line cap=round,
      line join=round,
    },
    blue edge/.style={
      draw=figBlue,
      line width=1.65pt,
      line cap=round,
      line join=round,
    },
    blue node shadow far/.style={
      circle,
      fill=figOutline,
      draw=none,
      opacity=0.05,
      minimum size=5.3mm,
      inner sep=0pt,
      transform canvas={xshift=1.0pt,yshift=-1.4pt},
    },
    blue node shadow near/.style={
      circle,
      fill=figOutline,
      draw=none,
      opacity=0.15,
      minimum size=4.3mm,
      inner sep=0pt,
      transform canvas={xshift=0.55pt,yshift=-0.8pt},
    },
    blue node outline/.style={
      circle,
      fill=figOutline,
      draw=none,
      minimum size=4.0mm,
      inner sep=0pt,
    },
    blue node/.style={
      circle,
      fill=figBlue,
      draw=none,
      minimum size=3.1mm,
      inner sep=0pt,
    },
    matching/.style={
      draw=figGray!78,
      dash pattern=on 2.2pt off 2.1pt,
      line width=1.05pt,
      line cap=round,
      shorten <=1.55mm,
      shorten >=1.55mm,
    },
    accent shadow far/.style={
      draw=figOutline,
      opacity=0.05,
      line width=5.2pt,
      line cap=round,
      line join=round,
      shorten <=1.0mm,
      shorten >=1.0mm,
      transform canvas={xshift=1.0pt,yshift=-1.4pt},
    },
    accent shadow near/.style={
      draw=figOutline,
      opacity=0.15,
      line width=4.0pt,
      line cap=round,
      line join=round,
      shorten <=1.05mm,
      shorten >=1.05mm,
      transform canvas={xshift=0.55pt,yshift=-0.8pt},
    },
    accent outline/.style={
      draw=figOutline,
      line width=3.35pt,
      line cap=round,
      line join=round,
      shorten <=1.15mm,
      shorten >=1.15mm,
    },
    removed/.style={
      draw=figOrange,
      line width=2.2pt,
      line cap=round,
      line join=round,
      shorten <=1.45mm,
      shorten >=1.45mm,
    },
    splice/.style={
      draw=figTeal,
      line width=2.2pt,
      line cap=round,
      line join=round,
      shorten <=1.45mm,
      shorten >=1.45mm,
    },
  ]
    \node[figure heading] at (3.22,3.72) {two cycles};
    \node[figure heading] at (10.78,3.72) {one circle};

    \foreach \x/\name in {1.05/t1,2.50/t2,3.95/t3,5.40/t4} {
      \coordinate (\name) at (\x,2.46);
    }
    \foreach \x/\name in {1.05/b1,2.50/b2,3.95/b3,5.40/b4} {
      \coordinate (\name) at (\x,1.10);
    }
    \draw[matching] (t1)--(b2) (t2)--(b1) (t3)--(b4) (t4)--(b3);
    \draw[accent shadow far] (t4)
      .. controls +(0,0.70) and +(0,0.70) .. (t1);
    \draw[accent shadow far] (b1)
      .. controls +(0,-0.70) and +(0,-0.70) .. (b4);
    \draw[accent shadow near] (t4)
      .. controls +(0,0.70) and +(0,0.70) .. (t1);
    \draw[accent shadow near] (b1)
      .. controls +(0,-0.70) and +(0,-0.70) .. (b4);
    \draw[accent outline] (t4)
      .. controls +(0,0.70) and +(0,0.70) .. (t1);
    \draw[accent outline] (b1)
      .. controls +(0,-0.70) and +(0,-0.70) .. (b4);
    \draw[removed] (t4)
      .. controls +(0,0.70) and +(0,0.70) .. (t1);
    \draw[removed] (b1)
      .. controls +(0,-0.70) and +(0,-0.70) .. (b4);
    \draw[blue shadow far] (t1)--(t2)--(t3)--(t4);
    \draw[blue shadow far] (b1)--(b2)--(b3)--(b4);
    \draw[blue shadow near] (t1)--(t2)--(t3)--(t4);
    \draw[blue shadow near] (b1)--(b2)--(b3)--(b4);
    \foreach \p in {t1,t2,t3,t4,b1,b2,b3,b4} {
      \node[blue node shadow far] at (\p) {};
      \node[blue node shadow near] at (\p) {};
    }
    \draw[blue outline] (t1)--(t2)--(t3)--(t4);
    \draw[blue outline] (b1)--(b2)--(b3)--(b4);
    \foreach \p in {t1,t2,t3,t4,b1,b2,b3,b4} {
      \node[blue node outline] at (\p) {};
    }
    \draw[blue edge] (t1)--(t2)--(t3)--(t4);
    \draw[blue edge] (b1)--(b2)--(b3)--(b4);
    \foreach \p in {t1,t2,t3,t4,b1,b2,b3,b4} {
      \node[blue node] at (\p) {};
    }

    \draw[figGray,line width=0.95pt,line cap=round,-{Stealth[length=2.3mm,width=1.6mm]}]
      (6.55,1.78)--(7.45,1.78);

    \foreach \x/\name in {8.60/u1,10.05/u2,11.50/u3,12.95/u4} {
      \coordinate (\name) at (\x,2.46);
    }
    \foreach \x/\name in {8.60/l1,10.05/l2,11.50/l3,12.95/l4} {
      \coordinate (\name) at (\x,1.10);
    }
    \draw[matching] (u1)--(l2) (u2)--(l1) (u3)--(l4) (u4)--(l3);
    \draw[accent shadow far] (u1)--(l1) (u4)--(l4);
    \draw[accent shadow near] (u1)--(l1) (u4)--(l4);
    \draw[accent outline] (u1)--(l1) (u4)--(l4);
    \draw[splice] (u1)--(l1) (u4)--(l4);
    \draw[blue shadow far] (u1)--(u2)--(u3)--(u4);
    \draw[blue shadow far] (l1)--(l2)--(l3)--(l4);
    \draw[blue shadow near] (u1)--(u2)--(u3)--(u4);
    \draw[blue shadow near] (l1)--(l2)--(l3)--(l4);
    \foreach \p in {u1,u2,u3,u4,l1,l2,l3,l4} {
      \node[blue node shadow far] at (\p) {};
      \node[blue node shadow near] at (\p) {};
    }
    \draw[blue outline] (u1)--(u2)--(u3)--(u4);
    \draw[blue outline] (l1)--(l2)--(l3)--(l4);
    \foreach \p in {u1,u2,u3,u4,l1,l2,l3,l4} {
      \node[blue node outline] at (\p) {};
    }
    \draw[blue edge] (u1)--(u2)--(u3)--(u4);
    \draw[blue edge] (l1)--(l2)--(l3)--(l4);
    \foreach \p in {u1,u2,u3,u4,l1,l2,l3,l4} {
      \node[blue node] at (\p) {};
    }

    \node[figure note,text=figOrange] at (3.22,-0.24)
      {delete the orange edges};
    \node[figure note,text=figTeal] at (10.78,-0.24)
      {add the green edges};
    \coordinate (wheel south west) at (current bounding box.south west);
    \coordinate (wheel north east) at (current bounding box.north east);
    \begin{scope}[on background layer]
      \node[soft panel,fit=(wheel south west)(wheel north east),inner sep=11pt] {};
    \end{scope}
  \end{tikzpicture}
  \caption{
    The wheel splice.
    Dashed edges belong to a matching and are not changed.
  }
  \label{fig:wheel-splice}
\end{figure}

The operation is illustrated in \cref{fig:wheel-splice}.

The splice preserves degrees, contact incidences, and matching equations.
Relabel the resulting cycle so that a contact is its last vertex, and designate
the edge from that contact to the first vertex as the circle-border equation.
This gives exactly the circle structure required below.  The only price is the
loss in the third conclusion of \cref{lem:wheel-splice}: after the splice, the
majority-rounding argument can increase the number of violations by at most two
per amplifier.  We account for this loss explicitly.

\paragraph{The $(1,2)$-\TSP black box.}
We use the construction and normalization of Karpinski and Schmied as a black
box~\cite[Theorem~6(i) and Section~7]{karpinski2012approximation}.  Suppose a
Hybrid instance of the form above has $m_2$ binary equations, $m_3$ original
three-variable equations, and $n_\circ$ circles.  Their construction produces
a complete $(1,2)$-\TSP instance on
\begin{equation}
  N=8m_2+27m_3+3n_\circ+1
  \label{eq:one-two-vertex-count}
\end{equation}
vertices, with the following two guarantees:
\begin{itemize}
  \item an assignment violating at most $u$ Hybrid equations yields a tour of
        length at most $N+u$; and
  \item a tour of length $N+u$ can be normalized and decoded into an assignment
        violating at most $u$ Hybrid equations.
\end{itemize}
The first term in \eqref{eq:one-two-vertex-count} counts eight vertices for each
of the $m_2+3m_3$ parity gadgets.  The chain joining the three-variable gadgets
contributes $3m_3+1$ vertices, and the circle-border paths contribute
$3n_\circ$ vertices.  This also explains why the base tour length is exactly
$N$.  More generally, every tour has length $N+u$, where $u$ is simply the
number of its length-two edges.

The construction and its normalization use only the one-cycle structure, the
perfect matching on checkers, and the incidences between contacts and original
equations.  The splice preserves these features; it also leaves every Hybrid
variable with exactly three occurrences and leaves every original equation in
the required three-variable parity form.  Thus the black-box guarantees apply
even though our circle spacing and total equation counts differ from the
particular Hybrid family used to state the original theorem.

\paragraph{Counting the instance and the splice loss.}
There are $3E$ occurrences of original variables and $d$ contacts per
amplifier, hence $3E/d$ amplifiers and circles.  An amplifier with $\rho d$
vertices has $\rho d$ circle equations and $(\rho-1)d/2$ matching equations.
Therefore
\[
\begin{aligned}
  m_2
  &=\frac{3E}{d}\left(\rho d+\frac{(\rho-1)d}{2}\right)
    =\frac{9\rho-3}{2}E,\\
  m_3&=E,
  &n_\circ&=\frac{3E}{d}.
\end{aligned}
\]
Substitution in \eqref{eq:one-two-vertex-count} gives
\[
  N=\left(36\rho+15+\frac9d\right)E+1.
\]
Set
\[
  c=36\rho+15+\frac9d,
  \qquad\text{so that}\qquad
  E=\frac{N-1}{c}.
\]

It remains to quantify the effect of the splice on soundness.  Fix one
amplifier and let $q$ be the number of contacts assigned the minority value.
Before the splice, the amplifier inequality gives at least $q$ violated
equality equations.  By \cref{lem:wheel-splice}, the splice can reduce this
number by at most two, so at least $\max\{q-2,0\}$ equalities remain violated.
Rounding the amplifier to its majority value repairs these equalities and can
newly violate at most the $q$ original equations incident to minority contacts.
The number of violations thus increases by at most
\[
  q-\max\{q-2,0\}=\min\{q,2\}\leq2
\]
per amplifier, and by at most
\[
  2\cdot\frac{3E}{d}=\frac{6E}{d}
\]
over the whole Hybrid instance.

\begin{proofof}{\cref{lem:one-two-gap}}
Fix $0<\xi<5/5268$ and write
\[
  c_0=36(69/10)+15=\frac{2634}{10}.
\]
Choose $0<\delta<\min\{1/4,15\xi\}$ and then $\eta>0$ sufficiently small that
\[
  \frac{1/2-\delta}{c_0+36\eta}>\frac5{2634}-\xi.
\]
Choose $d$ from the unbounded amplifier family sufficiently large that the
fixed-occurrence hardness holds and
\[
  \frac{1/2-\delta-6/d}{c_0+36\eta+9/d}
  \geq\frac5{2634}-\xi.
\]
Finally, fix an amplifier with density $\rho\leq69/10+\eta$.  Since $c\geq15$
and $c\leq c_0+36\eta+9/d$, these choices give
\[
  \frac{\delta}{c}\leq\xi
  \qquad\text{and}\qquad
  \frac{1/2-\delta-6/d}{c}\geq\frac5{2634}-\xi.
\]

In a YES instance, assign every vertex of an amplifier the value of its
original variable.  All equality equations are satisfied, so an assignment
violating at most $\delta E$ original equations gives a tour of length at most
\[
  N+\delta E
  =N+\frac{\delta}{c}(N-1)
  \leq(1+\xi)(N-1)+1.
\]

Conversely, suppose that a tour has length $N+u$.  The Karpinski--Schmied
normalization gives a Hybrid assignment with at most $u$ violations.  Majority
rounding gives an assignment to the original variables with at most
$u+6E/d$ violated original equations.  In a NO instance this is at least
$(1/2-\delta)E$, and therefore
\[
\begin{aligned}
  u
  &\geq\left(\frac12-\delta-\frac6d\right)E\\
  &=\frac{1/2-\delta-6/d}{c}(N-1)
   \geq\left(\frac5{2634}-\xi\right)(N-1).
\end{aligned}
\]
Thus every NO-instance tour has length at least
$\left(1+5/2634-\xi\right)(N-1)+1$, which is the claimed gap.  The constants
$d$, $\rho$, and the chosen amplifier depend only on $\xi$ and are hardwired
into the construction, so the reduction is polynomial-time.
\end{proofof}

\end{document}